\documentclass[letterpaper,twocolumn,american,prl,superscriptaddress]{revtex4-1}
\usepackage[latin9]{inputenc}
\usepackage{amsmath}
\usepackage{graphicx}
\usepackage{esint}

\makeatletter

\usepackage{babel}
\begin{document}

\title{Spontaneous charge carrier localization in extended one-dimensional
systems}

\author{Vojt\v{e}ch Vl\v{c}ek}

\affiliation{Fritz Haber Center for Molecular Dynamics, Institute of Chemistry,
The Hebrew University of Jerusalem, Jerusalem 91904, Israel}

\affiliation{Bayerisches Geoinstitut, Universität Bayreuth, 95440 Bayreuth, Germany}

\author{Helen R. Eisenberg}

\affiliation{Fritz Haber Center for Molecular Dynamics, Institute of Chemistry,
The Hebrew University of Jerusalem, Jerusalem 91904, Israel}

\author{Gerd Steinle-Neumann}

\affiliation{Bayerisches Geoinstitut, Universität Bayreuth, 95440 Bayreuth, Germany}

\author{Daniel Neuhauser}

\affiliation{Department of Chemistry and Biochemistry, University of California,
Los Angeles California 90095, U.S.A.}

\author{Eran Rabani}

\affiliation{Department of Chemistry, University of California and Materials Science
Division, Lawrence Berkeley National Laboratory, Berkeley, California
94720, U.S.A.}

\affiliation{The Sackler Center for Computational Molecular and Materials Science,
Tel Aviv University, Tel Aviv, Israel 69978}

\author{Roi Baer}

\affiliation{Fritz Haber Center for Molecular Dynamics, Institute of Chemistry,
The Hebrew University of Jerusalem, Jerusalem 91904, Israel}

\altaffiliation{On sabbatical in the Department of Chemistry, University of California, Berkeley California 94720, U.S.A.}

\email{roi.baer@huji.ac.il}

\date{\today}
\begin{abstract}
Charge carrier localization in extended atomic systems has been described
previously as being driven by disorder, point defects or distortions
of the ionic lattice. Here we show for the first time by means of
first-principles computations that charge carriers can spontaneously
localize due to a purely electronic effect in otherwise perfectly
ordered structures. Optimally-tuned range-separated density functional
theory and many-body perturbation calculations within the GW approximation
reveal that in trans-polyacetylene and polythiophene the hole density
localizes on a length scale of several nanometers. This is due to
exchange-induced translational symmetry breaking of the charge density.
Ionization potentials, optical absorption peaks, excitonic binding
energies and the optimally-tuned range parameter itself all become
independent of polymer length as it exceeds the critical localization
scale. Moreover, lattice disorder and the formation of a polaron result
from the charge localization in contrast to the traditional view that
lattice distortions precede charge localization. Our results can explain
experimental findings that polarons in conjugated polymers form instantaneously
after exposure to ultrafast light pulses. 
\end{abstract}


\maketitle
Spatial localization in extended systems has been a central topic
in physics, since the pioneering work of Anderson~\cite{Anderson1958}
and Mott~\cite{Mott1968}, and more recently in the context of many-body
localization~\cite{Basko2006}. It also forms an important theme
in materials science of extended conjugated systems where the dynamics
of charges carrier are described in terms of localized polarons.~\cite{Friend1999,Brandes2003,Scholes2006,Johns2010,McMahon2010,Lannoo2012,Noriega2013}.
One way to identify charge localization is through the dependence
of its energy (e.g., ionization potential or electron affinity) on
the system size $L$. In 1D systems, if the charge remains delocalized,
then according to a simple \emph{non-interacting} picture, its energy
converges to the bulk limit as $1/L^{\alpha}$ with $\alpha=1$ for
a metal or $\alpha=2$ otherwise. However, if the energy becomes independent
of $L$ for $L>\ell_{c}$, it could be due to charge localization
within a critical length scale~$\ell_{c}$. 

Charge localization in conjugated systems can occur in several ways:
Attachment by point defects~\cite{Lannoo2012}, lattice disorder
effects~\cite{Brandes2003,Noriega2013}, and formation of self-bound
charged polarons and neutral solitons by local distortion of the nuclear
lattice~\cite{Kurlancheek2012,Salzner2014,Korzdorfer2014,Kohler2015}.
However, it still remains an open question whether localization can
occur in disorder-free transitionally invariant systems. This question
has received much attention recently in the context of many-body localization~\cite{Yao2014,DeRoeck2014,Hickey2014,Schiulaz2015}. 

In this letter we provide first-principles computational evidence
for a new mechanism of localization in 1D conjugated systems, in which
the electrons form their own nucleation center without the need to
introduce disorder into the Hamiltonian. This challenges the widely
accepted picture in which the electronic eigenstates localize only
after coupling with the lattice distortion~\cite{Holstein1959}.
To illustrate this mechanism, we study the electronic structure and
the charge distribution in large one-dimensional systems with ideal
(ordered) geometries. We focus on two representative conjugated polymers,
trans-polyacetylene (tPA) and polythiophene (PT), with increasing
lengths $L=M\ell_{1}$ up to $M=70$ and $M=20$, respectively ($\ell_{1}$
is the length of the repeat unit). Besides their practical significance~\cite{Scholes2006},
they also exhibit interesting physical phenomena, in which polarons,
bipolarons and solitons affect charge mobility and localization~\cite{Puschnig2003,Friend1999,Nayyar2011,Hoffmann2013,Salzner2014}.

\begin{figure}[t]
\includegraphics[width=8cm]{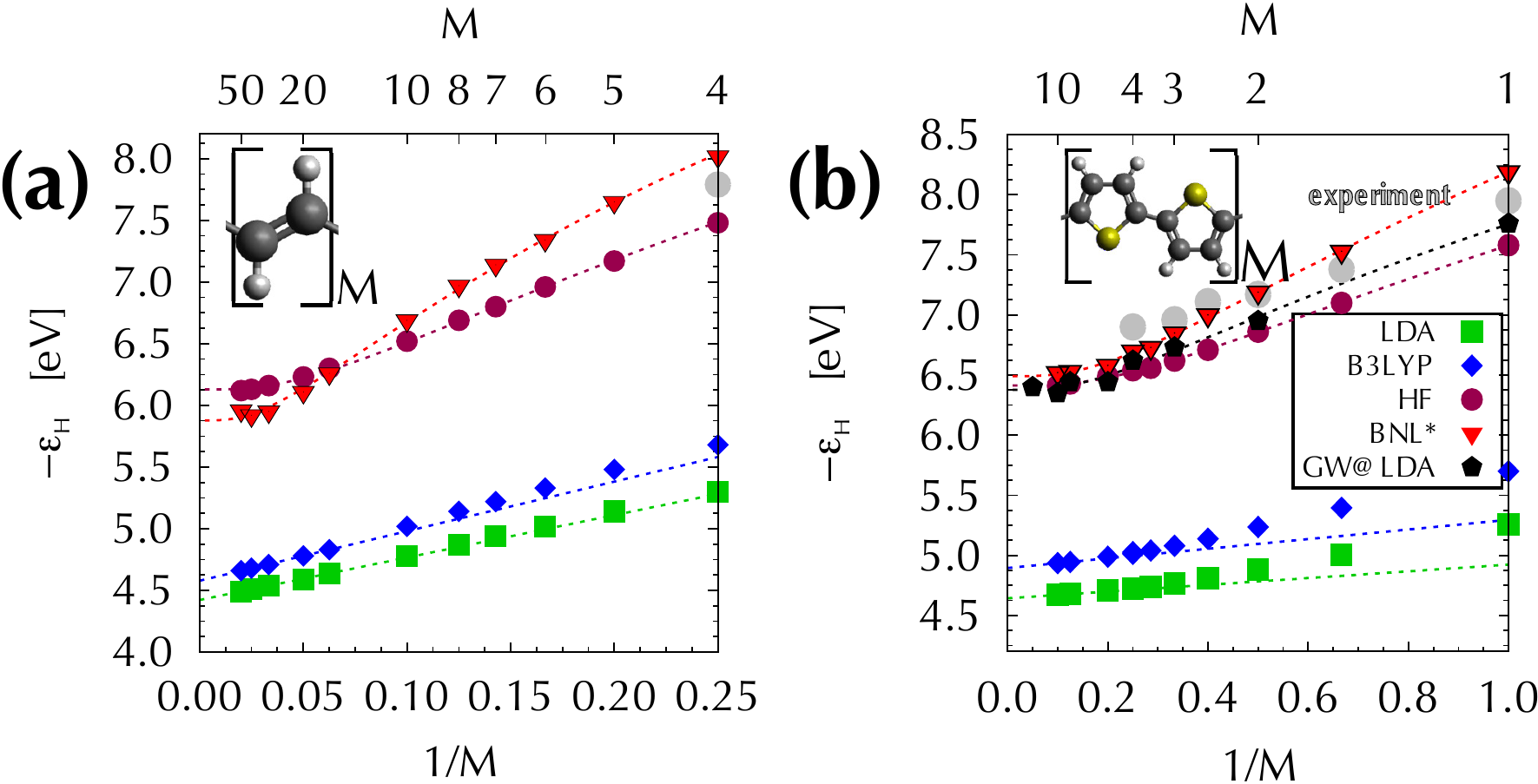} \caption{\label{fig:figip}Ionization potentials (estimated using highest occupied
eigen-energies $\varepsilon_{H}$) for \textbf{(a)} trans-polyacetylene
and \textbf{(b)} polythiophene shown against the inverse number of
repeat units $M$ in the respective polymer. The repeat unit for each
polymer is illustrated in the corresponding insets (C, H and S are
shown by black, white and yellow spheres, respectively). Results obtained
from different computational approaches are indicated by colors and
labelled in the figure. Experimental data for the ionization potentials
(gray circles) were taken from Refs.~\citenum{Jones1990,Filho2007,PinheiroArXiv}
and references therein. The dashed lines represent a numerical fit
to $-\varepsilon_{H}\left(M\right)=-\varepsilon_{H}\left(\infty\right)+\frac{\Delta\varepsilon}{M}$
for LDA and B3LYP ($\varepsilon_{H}\left(\infty\right)$ and $\Delta\varepsilon$
are fitting parameters) and to $-\varepsilon_{H}\left(M\right)=-\varepsilon_{H}\left(\infty\right)+\Delta\varepsilon\exp\left(-\sqrt{M/M_{0}}\right)$
for HF, BNL{*}, and GW.}
\end{figure}

\begin{figure}[t]
\includegraphics[width=8cm]{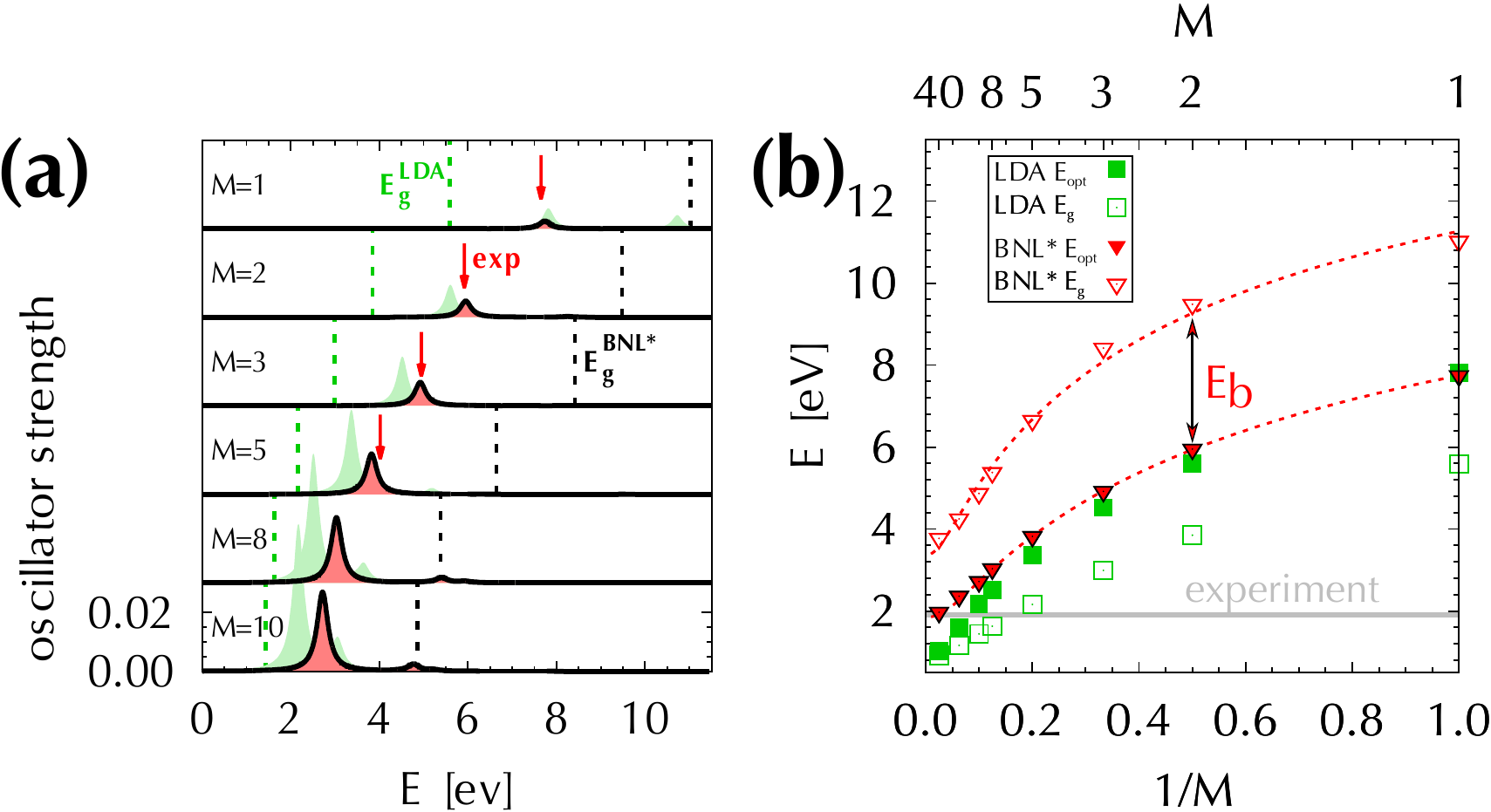} \caption{\label{fig:optical} \textbf{(a)} Calculated optical spectra for selected
tPA polymers of various lengths (numbers of repeat units $M$). All
calculations were performed with the cc-pvTZ basis set using TDDFT
within the BNL{*} functional (solid black line with red fill) and
LDA functional (green filled curve). The fundamental band gaps are
shown by dashed vertical lines in corresponding colors. Red arrows
indicate experimental absorption peak positions (Refs.~\citenum{McDiarmid1976,Flicker1977,Damico1980,Bredas1983}
and references therein). \textbf{(b)} Position of the first maxima
of the absorption $E_{opt}$ and the fundamental band gap $E_{g}$
obtained with BNL{*} and LDA functionals as function of inverse number
of repeat units. Results for the two longest polymers were calculated
using the 3-21G basis set, other results are obtained using cc-pvTZ.
The exciton binding energy is the difference between $E_{g}$ and
the peak maximum is illustrated by an arrow. The horizontal full line
represents the experimental energy of the maximum absorption for the
infinite system (1.9 eV)~\cite{Feldblum1982}. }
\end{figure}

In Fig.~\ref{fig:figip} we plot the ionization potentials (IPs)
for both the tPA ( panel a) and PT (panel b) polymers as a function
of the number of repeat units, $M$. To illustrate the effect of localization
we focus on the ionization potential representing the energy of positive
charge carrier (hole) rather than on the electron affinity representing
the energy of negative charge carrier (electron), since we find the
former to localize on shorter length scales (see below). Several levels
of theory are used: Hartree-Fock (HF) theory, density functional theory
(DFT) within the local density approximation (LDA)~\cite{PerdewWang},
the optimally-tuned BNL{*}~\cite{BaerNeuhauser2005,Baer2010,Kronik2012}
range-separated hybrid functional~\cite{Savin1995}, and the B3LYP~\cite{B3LYP}
approximation, and, finally, the $\mbox{G}_{0}\mbox{W}_{0}$ many-body
perturbation technique~\cite{HybertsenLouie} (on top of of LDA implemented
using stochastic DFT \cite{Baer2013}) within the stochastic formulation
(\emph{s}GW)~\cite{Neuhauser2014}. The LDA and to some extent the
B3LYP lack sufficient exact exchange while HF lacks correlations and
screening effects. BNL{*} provide a systematic description of correlations
and exact exchange through the process of optimal tuning~\cite{Livshits2007}.
$\mbox{G}_{0}\mbox{W}_{0}$ is based on many-body perturbation theory
and includes exchange, correlation and screening effects~\cite{HybertsenLouie}.

The LDA and B3LYP computations yield IPs that are considerably smaller
than the experimental values, consistent with previous computational
studies on shorter polymer chains~\cite{Salzner2010,Salzner2011}
and with general theoretical arguments~\cite{Salzner2009,Stein2012}.
These IP values drop to their bulk limit ($I_{\infty}^{tPA}=4.4$~eV
for tPA and $I_{\infty}^{PT}=4.6$~eV for PT) asymptotically linearly
as $M^{-1}$ for the range of sizes studied (they do not fit the purely
non-interacting asymptotic dependence of $M^{-2}$). In contrast,
HF IPs are significantly closer to the experimental values, deviating
by less than $0.4$~eV. The HF IPs also initially drop as polymer
size increases but for a polymer of length exceeding a critical value,
they quickly converge to an asymptotic value, hinting at localization
of the hole. The asymptotic HF IPs and HF critical length scale are
$I_{\infty}^{tPA}=6.1$~eV and $\ell_{c}^{tPA}=4.9$~nm and $I_{\infty}^{PT}=6.4$~eV
and $\ell_{c}^{PT}=3.1$~nm (see Supplementary Material for the approach
used to determine these quantities). The computational IPs of BNL{*}
and the \emph{s}GW are in even better agreement with the available
experimental data than those of HF. They too show a localization transition
with $I_{\infty}^{tPA}=5.9$~eV and $\ell_{c}^{tPA}=7.9$~nm for
tPA and $I_{\infty}^{PT}=6.4-6.5$~eV and $\ell_{c}^{PT}=4.2-4.3$~nm
for PT. Using the results for intermediate polymers (which do not
exhibit localization yet) we can linearly extrapolate to the limit
$M\to\infty$ and estimate the value of ionization potential if no
localization occurs; this yields IP values smaller by $\approx0.5$~eV
which can be viewed as the energy of spontaneous localization. While
the asymptotic values of the ionization potentials predicted by HF,
BNL{*} and \emph{s}GW are similar, the BNL{*} and \emph{s}GW critical
length scales $\ell_{c}$ are larger than those predicted by HF. This
result is consistent with the tendency of HF to over-localize holes
in finite systems~\cite{MoriSanchez2008,Livshits2008}. 

\begin{figure*}[!tbph]
\centering{}\includegraphics[width=17cm]{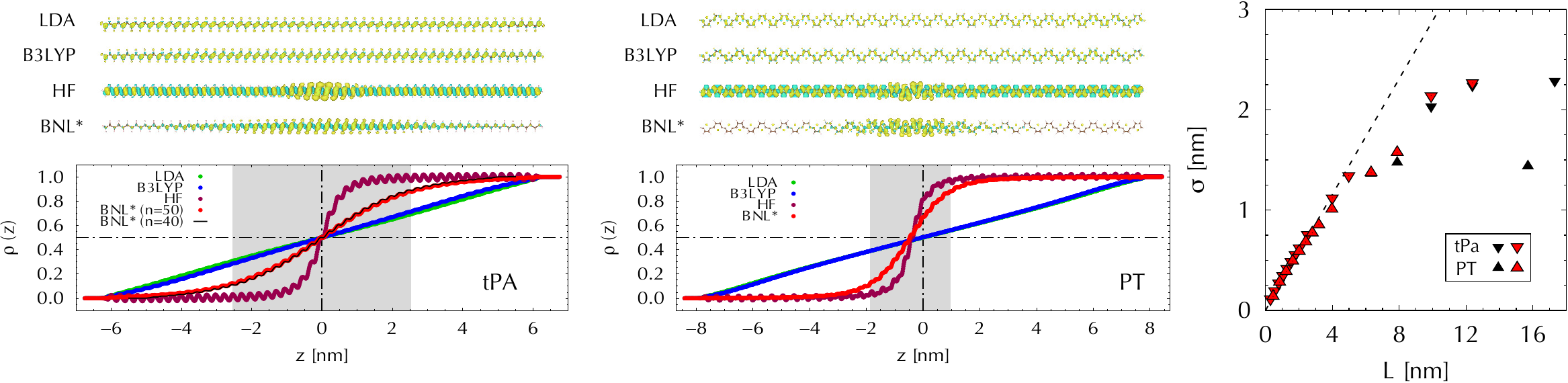}
\caption{\label{fig:hole-density}Left panels: The hole densities (top two
panels), $\Delta n\left(\mathbf{r}\right)$, for the corresponding
labelled methods in long strands of $M=50$ repeat tPA units (left)
and $M=20$ repeat PT units (right). The hole is shown as a yellow
(aqua) 0.00025$a_{0}^{-3}$ (-0.00025$a_{0}^{-3}$) density isosurface.
In the two bottom left panels we plot the cumulative density, $\rho\left(z\right)$,
for different functionals. The cumulative curve for a shorter tPA
polymer with $M=40$ (black line) is practically indistinguishable
from $M=50$. Gray areas in the plots show the value of the second
cumulant ($\sigma$) for the corresponding BNL{*} hole density, which
are plotted in the right panel for different polymer lengths. The
dashed straight line in the right panel is the fully delocalized result
($\sigma=L/\sqrt{12}$). Note that for the larger system we used a
smaller basis (3-21G, black symbols) which closely follow the results
using a larger basis (cc-pvTZ, red symbols).}
\end{figure*}

To further strengthen the validity of the BNL{*} treatment (and indirectly
the $\mbox{G}_{0}\mbox{W}_{0}$ which agrees with the BNL{*}), we
compare its predicted optical excitations $E_{opt}$ and \emph{fundamental
gaps }$E_{g}=\varepsilon_{L}-\varepsilon_{H}$ in tPA to experimental
results, where available~\cite{McDiarmid1976,Flicker1977,Damico1980,Bredas1983}
(see Table II of the Supplementary material). The absorption spectra
shown in the left panel of Fig.~\ref{fig:optical} were calculated
using time-dependent LDA (ALDA) and BNL{*} (ABNL{*}) functionals~\cite{Stein2009,Baer2010}.
It is seen that the ABNL{*} approach provides excellent agreement
for the optical gaps $E_{opt}^{ABNL*}$ in comparison to experimental
data. The optical gaps $E_{opt}$ are also plotted as a function of
$1/M$ on the right panel of Fig.~\ref{fig:optical} and it is seen
that for the largest system studied the ABNL{*} optical gap is in
excellent agreement with the experimental value~\cite{Feldblum1982,Leising1988}.
This is in contrast to the ALDA results which underestimate this limit
by $\approx1$~eV, and consistently deviate from the ABNL{*} results
as the system size increases. In the right panel of Fig.~\ref{fig:optical}
we also plot the fundamental gap $E_{g}^{BNL*}$. The values of $E_{g}^{BNL*}$
for small systems yields excellent agreement with previous $\mbox{G}_{0}\mbox{W}_{0}$
results~\cite{PinheiroArXiv}. Furthermore, $E_{g}^{BNL*}$ does
not localize for the tPA lengths studied. Since, $\varepsilon_{H}^{BNL*}$
localizes within a length scale of $\ell_{c}=7.9$~nm the continued
change in $E_{g}^{BNL*}$ for larger polymers must result from a continued
change in the electron energy $\varepsilon_{L}^{BNL*}$. This suggests
that negative added charge does not yet localize for the tPA sizes
studied and may explain why the finite size gaps are larger than the $\mbox{G}_{0}\mbox{W}_{0}$
gap of 2.1 eV for $L\rightarrow\infty$ \cite{Rohlfing1999,Puschnig2003,Gorling2006}.
Note, however, that the $\mbox{G}_{0}\mbox{W}_{0}$ gaps are rather
sensitive to the size of the unit cell and small changes of 0.005
nm in the position of the atoms can lead to significant fluctuation
of $2.0$ to $4.2$~eV in the gaps~\cite{Ferretti2012}. Since there
are no experimental measurements of the fundamental gap when $L\rightarrow\infty$,
it still remains an open question as to the length scale at which
\emph{electrons} localize (as opposed to hole localization, which
already occurs at the system sizes studied). To reach system sizes
at which the electron localizes will probably require using a stochastic
approach for BNL{*}~\cite{Neuhauser2015}. Finally, panel a of Fig.~\ref{fig:optical}
shows that the exciton binding energy $E_{b}=E_{g}-E_{opt}$ is on
the order of $E_{g}/2$ for the larger systems, a value typical of
other 1D conjugated systems~\cite{Wang2005}, indicating that neutral
excitations are dominated by electron-hole interactions.

Up to now we have studied localization only from the point of view
of energy changes. It is instructive to also study localization in
terms of the \emph{hole density, }which is the difference $\Delta n\left(\mathbf{r}\right)=n^{N}\left(\mathbf{r}\right)-n^{N-1}\left(\mathbf{r}\right)$
between the ground state density of the neutral ($N$) and the ground
state density of the positively charged ($N-1$) systems. For non-interacting
electrons this quantity equals the density of the highest occupied
eigenstate, which is not localized. However, for interacting electrons
$\Delta n\left(\mathbf{r}\right)$ must be calculated as the difference
of densities obtained from two \emph{separate }self-consistent field
DFT calculations and can thus exhibits a different behavior. We have
also ascertained that the same localization pattern emerges even when
an infinitesimal charge $q\to0$ is removed, showing that localization
of the hole density occurs even also in the linear response regime.

The isosurface plots of the hole densities are given in the upper
left and middle panels of Fig.~\ref{fig:hole-density} for the various
methods (excluding \emph{s}GW). In the lower left and middle panels
we show the \emph{cumulative} hole densities $\rho\left(z\right)=\int_{-\infty}^{z}{\rm d}z^{\prime}\int_{-\infty}^{\infty}{\rm d}y^{\prime}\int_{-\infty}^{\infty}{\rm d}x^{\prime}\Delta n\left(\mathbf{r^{\prime}}\right)$.
In both types of plots it is evident that LDA and B3LYP do not show
localization of the hole density in any of the systems studied and
in $\rho\left(z\right)$ they show a linear monotonic increase. Contrarily,
the HF and BNL{*} charge distributions localize as observed by change
of $\rho\left(z\right)$ near the center of the chain. In PT this
transition in $\rho\left(z\right)$ occurs around one of the S atoms
closest to the center of the polymer, due to the lack of mirror plane
symmetry. For long polymers exceeding $\ell_{c}$, the BNL{*} hole
density hardly changes as seen by the overlapping $\rho\left(z\right)$
of polymers with $M=40$ and $M=50$. This implies that the size of
the hole is no longer influenced by the polymer terminal points and
is thus independent of system size. 

The extent of hole localization can be described by the second cumulant
$\sigma=\sqrt{\int\Delta n\left(\mathbf{r}'\right)\left(z'-\bar{z}\right)^{2}d\mathbf{r}'}$
(where $\bar{z}=\int\Delta n\left(\mathbf{r}'\right)z'd\mathbf{r}'$).
This is shown in the right panel of Fig.~\ref{fig:hole-density}
for BNL{*}. For small sizes $\sigma$ increases as $L/\sqrt{12}$,
consistent with a uniform hole density spread over the entire polymer.
As $L$ increases beyond $\ell_{c}$, the BNL{*} $\sigma$ converge
to an asymptotic value, $\sigma_{\infty}$, while those of LDA continue
to follow the linear $L/\sqrt{12}$ law (not shown). 

It is important to note that the hole density $\Delta n\left(\mathbf{r}\right)$
is dominated by the minority-spin density changes: the orbitals having
the same spin as the removed electron redistribute such as to localize
the hole density near the chain center. On the other hand, the majority-spin
orbitals remain nearly unperturbed and thus do not contribute to $\Delta n\left(\mathbf{r}\right)$.
This fact reveals that the localization is driven by attractive non-local
exchange interactions existing solely between like-spin electrons.
This is further supported by the fact that localization only appears
in methods that account for non-local exchange (HF, BNL{*}, and $\mbox{G}_{0}\mbox{W}_{0}$). 

One of the interesting ramifications of the IP stabilization as polymer
length exceeds a critical length scale is the simultaneous stabilization
of the BNL{*} range-separation parameter $\gamma$. This is because
in the absence of hole localization the tuning criterion,\cite{Livshits2007}
$I+\varepsilon_{H}=0$ is expected to become automatically satisfied
when (semi)local functionals are used in the limit of infinite system
size~\cite{GodbyWhite,Ogut1997,MoriSanchez2008,Vlcek2015} forcing
$\gamma$ (and with it the non-local exchange part of the functional)
to drop eventually to zero. In the systems studied here localization
saves the day for tuning and the range parameter attains finite asymptotic
values of $\gamma^{tPA}=2.7$~nm$^{-1}$ and $\gamma^{PT}=3.1$~nm$^{-1}$.
The leveling of $\gamma$ with $L$ was reported for PT~\cite{Korzdorfer2011},
however, it was not previously clear whether $\gamma$ would level-off
for tPA. It is worth pointing out that $\gamma$ does not change significantly
($<0.2\,\mbox{nm}^{-1}$) when LYP correlation is used instead of
LDA in the BNL{*} calculation.

\begin{figure}[t]
\begin{centering}
\includegraphics[width=8.5cm]{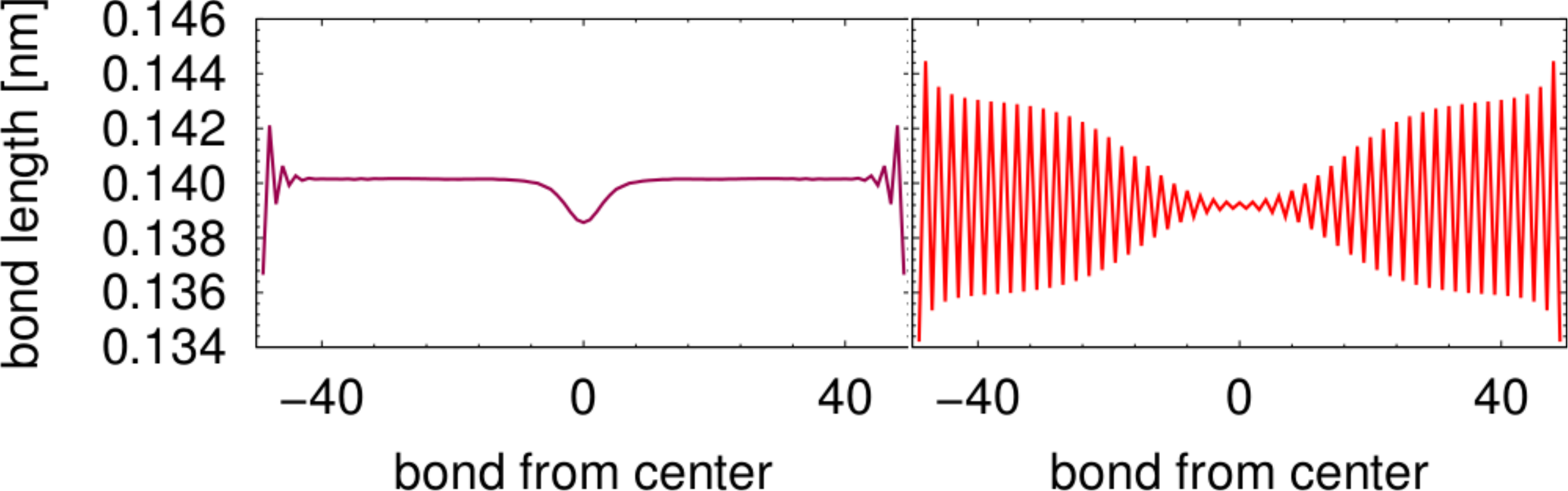}
\par\end{centering}

\caption{\label{fig:C-C BLA}The C-C bond length in the charged $M=50$ tPA
polymer as predicted by HF (left panel) and BNL{*} (right panel) obtained
with the 3-21G basis set. In BNL{*}, a polaron appears as a reduction
of the bond-length alternation, while in the region about 40 C-C bonds
away from the polaron, the alternation is increased to $0.007$~nm,
similar to the experimental value of $0.008\,\mbox{nm}$ for neutral
chains~\cite{Yannoni1983}.}
\end{figure}

While HF supports partial localization, its hole density also exhibits
oscillations\emph{ }along the entire polymer length that do not diminish
with system size. These indicate a rigid shift of charge between neighboring
atoms: From double to single C-C bonds in tPA and from S to nearby
C atoms for PT. This is consistent with the tendency of HF to eliminate
bond-length alternation in the entire tPA polymer chain~\cite{Rodriguez-Monge1995}
as shown in the left panel of Fig.~\ref{fig:C-C BLA}. BNL{*} on
the other hand eliminates the bond-length alternation only in proximity
of the localized hole density (right panel of Fig.~\ref{fig:C-C BLA}),
consistent with a localized polaron model. 

In summary, using first principles density functional theory and many-body
perturbation theory, we have shown that positive charge carriers can
localize in 1D conjugated polymers due to a spontaneous, purely electronic
symmetry breaking transition. In this case, localization is driven
by non-local exchange interactions and thus cannot occur when (semi)local
density functional approximations are used. HF theory, which has non-local
exchange, shows a localization transition in a relatively small length-scale
but predicts complete annihilation of bond-length alternation upon
ionization, irrespective of polymer length. BNL{*}, which through
tuning includes a balanced account of local and non-local exchange
effects, provides an accurate description of the optical gap in comparison
to experiments and shows a localization transition with a length scale
(estimated from the leveling off of the IPs) that agrees well with
the \emph{s}GW approach. Moreover, BNL{*} predicts a localized disruption
of the bond-length alternation. 

The localization phenomenon is driven by the same-spin attractive
non-local exchange interactions and therefore, cannot be explained
in terms of classical electrostatics. There is no reason to assume
that the observed emergence of the localization length $\ell_{c}$
in finite systems will not readily occur also in infinite systems,
where hole states near the top of the valence band are necessarily
infinitely degenerate.

We thank professors Ulrike Salzner and Leeor Kronik for illuminating
discussions on polymers and localization in large systems. R.B. and
E.R. are supported by The Israel Science Foundation--FIRST Program
(Grant No. 1700/14). V.V. is supported by Minerva Stiftung of the
Max Planck Society, R.B. gratefully acknowledges support for his sabbatical
visit by the Pitzer Center and the Kavli Institute of the University
of California, Berkeley. D.N. and E.R. acknowledge support by the
NSF, grants CHE-1112500 and CHE-1465064, respectively. This research
used resources of the National Energy Research Scientific Computing
Center, a DOE Office of Science User Facility supported by the Office
of Science of the U.S. Department of Energy under Contract No. DE-AC02-05CH11231,
and at the Leibniz Supercomputing Center of the Bavarian Academy of
Sciences and the Humanities.

\bibliographystyle{apsrev4-1}
\bibliography{library}

\end{document}